\RequirePackage{ifpdf}
\ifpdf 
\documentclass[pdftex]{sigma}
\else
\documentclass{sigma}
\fi

\begin{document}
\allowdisplaybreaks
\renewcommand{\PaperNumber}{019}

\FirstPageHeading

\ShortArticleName{Transverse Evolution Operator for the
Gross--Pitaevskii Equation}

\ArticleName{Transverse Evolution Operator\\
for the Gross--Pitaevskii Equation\\ in Semiclassical
Approximation}

\Author{Alexey BORISOV~$^{\dag}$, Alexander
SHAPOVALOV~$^{\dag\ddag\S}$ and Andrey TRIFONOV~$^{\ddag\S}$}

\AuthorNameForHeading{A. Borisov, A. Shapovalov  and A. Trifonov}

\Address{$^\dag$~Tomsk State University, 36 Lenin Ave., 634050
Tomsk, Russia}
\EmailD{\href{mailto:borisov@phys.tsu.ru}{borisov@phys.tsu.ru},
\href{mailto:shpv@phys.tsu.ru}{shpv@phys.tsu.ru}}

\Address{$^\ddag$~Tomsk Polytechnic University, 30 Lenin Ave.,
634050 Tomsk, Russia}
\EmailD{\href{mailto:trifonov@mph.phtd.tpu.edu.ru}{trifonov@mph.phtd.tpu.edu.ru}}

\Address{$^\S$~Math. Phys. Laboratory, Tomsk Polytechnic
University, 30 Lenin Ave.,  634050 Tomsk, Russia}

\ArticleDates{Received July 27, 2005, in final form November 13,
2005; Published online November 22, 2005}

\Abstract{The Gross--Pitaevskii equation with a local cubic
nonlinearity that describes a~many-dimensional system in an
external field is considered in the framework of the complex
WKB--Maslov method. Analytic asympto\-tic solutions are
constructed in semiclassical approximation in a small parameter
$\hbar$, $\hbar\to 0$, in the class of functions concentrated in
the neighborhood of an unclosed surface associated with the phase
curve that describes the evolution of  surface vertex. The
functions of this class are of the  one-soliton form along the
direction of the surface normal. The general constructions are
illustrated by examples.}


\Classification{81Q20; 81R30; 35Q55}

\section{Introduction}

The Gross--Pitaevskii equation (GPE), derived independently by
Gross \cite{borisov:GROSS} and Pitaevskii
\cite{borisov:PITAEVSKII}, arises in various models of nonlinear
physical phenomena. This is a Schr\"odinger-type equation with an
external field potential $U(\vec x,t)$  and a local cubic
nonlinearity:
\begin{gather}
\label{borisov:GPE-1}  \left( -i\hbar
\partial _{t} + \frac{{\widehat{\vec p}}\, ^2}{2m}+ U(\vec r)+
\varkappa  |\Psi (\vec x,t,\hbar)|^{2} \right)\Psi (\vec
x,t,\hbar) =0.
\end{gather}
Here  $\vec x=(x_j)\in \mathbb{R}^n$; $i,j,k,l=1,\ldots,n$; $t\in
\mathbb{R}^1$; $\partial_t=\partial /\partial t$; $\widehat{\vec
p}=-i\hbar \nabla$; $\nabla $ is a gradient operator in  $\vec x$;
$\varkappa$ and $\hbar$  are  real parameters. The complex
function $\Psi(\vec x,t,\hbar)$ determines the system state,
$|\Psi|^2=|\Psi (\vec x,t,\hbar)|^2=\Psi (\vec x,t,\hbar)\Psi^{*}
(\vec x,t,\hbar)$; the function  $\Psi^{*} (\vec x,t,\hbar)$ is
complex conjugate to  $\Psi (\vec x,t,\hbar)$.

For $U=0$, $\varkappa <0$, and  $n=1$,
equation~(\ref{borisov:GPE-1}) is known to be integrable by the
Inverse Scattering Transform (IST) method \cite{borisov:ZAKHAROV,
borisov:ZAKHAROV-1} and to have exact soliton  solutions. The
solutions  are written in terms of space localized functions which
conserve their form during the evolution. Solitons have numerous
physical applications; for example, they serve  to describe the
propagation of optical pulse in nonlinear media
\cite{borisov:HASEGAWA,borisov:MOLLENAUER}.

The solutions of equation~(\ref{borisov:GPE-1}) that  are not
completely localized are also of interest, in parti\-cu\-lar, in
the theory of nonlinear deep-water waves where the two-dimensional
($n=2$) solutions of plane wave  solitonlike type are localized
only along the wave propagation direction \cite{borisov:YUENLAKE}.

The  Gross--Pitaevskii equation (\ref{borisov:GPE-1})  in physical
dimensions ($n=2,3$) is used in the mean-field quantum theory of
Bose-Einstein condensate (BEC) formed by ultracold  bosonic
coherent atomic ensembles~\cite{borisov:PITAEVSKII-1}. As a rule,
the function  $U(\vec x,t)$ is  the potential of an external field
of a~magnetic trap and laser radiation. The wave function $\Psi
(\vec x,t,\hbar)$ corresponds to a condensate state. The local
nonlinearity term $\varkappa |\Psi (\vec x,t,\hbar)|^{2}$ arises
from an assumption about the delta-shape interatomic potential.
For multidimensional cases, no method of exact integration of the
 GPE  is known, to our knowledge, and  equation~(\ref{borisov:GPE-1})
is usually studied for some special  cases by computer simulation.
However, this approach has natural restrictions and also has
nontrivial specific features as long as  localized solutions of
equation~(\ref{borisov:GPE-1}) are known
 to be unstable and to collapse when  $n>1$,
$U=0$, and $\varkappa < 0$ \cite{borisov:ZAKHAROV-2}. Therefore,
to develop  methods for constructing analytical solutions of the
GPE (\ref{borisov:GPE-1}) for many-dimensional cases is critical.

In this paper we construct a class of  asymptotic solutions for
the $(1+n)$-dimensional GPE  with a  focusing local cubic
nonlinearity
\begin{gather}
\label{borisov:GPE}  \widehat{L}(\Psi)(\vec x)=\left[ -i\hbar
\partial _{t} + {\mathcal H}(\hat {\vec p},\vec x,t) - g^{2}|\Psi
(\vec x,t,\hbar)|^{2} \right]\Psi (\vec x,t,\hbar) =0.
\end{gather}
Here $g$ is a real nonlinearity parameter and  $\hbar$ is an
asymptotic parameter, $\hbar \to 0$. The  pseudo-differential
operator ${\mathcal H}(\hat {\vec p},\vec x,t)$ is Weyl-ordered
\cite{borisov:KARASEV-MASLOV} and its symbol ${\mathcal H}( {\vec
p},\vec x,t)$ is quadratic in $\vec p\in \mathbb{R}^n$
\begin{eqnarray}
&&{\mathcal H}(\hat {\vec p},\vec x,t)=\frac{1}{2}\langle\hat
{\vec p}, {\mathcal H}_{pp}(t)\hat {\vec
p}\rangle+\frac12\left(\langle\hat {\vec p}, \vec{\mathcal H}(\vec
x,t)\rangle +\langle\vec{\mathcal H} (\vec x,t), \hat {\vec
p}\rangle\right)+ {\mathcal H}_{0}(\vec x,t), \label{borisov:GPE1}
\end{eqnarray}
where  $\langle \vec a,\vec b\rangle=$
$\sum\limits_{l=1}^{n}a_lb_l$; the real functions --- the
$(n\times n)$-matrix ${\mathcal H}_{pp}(t)$, the  vector
$\vec{\mathcal H} (\vec x,t)$, and  the scalar  $ {\mathcal
H}_{0}(\vec x,t)$ --- smoothly depend on their arguments. These
functions model the external fields imposed on the system
described by equation~(\ref{borisov:GPE}).

A formalism of semiclassical asymptotics was  developed for the
GPE with a nonlocal nonlinearity in
\cite{borisov:BTS,borisov:LTS}. This equation was named the
Hartree type equation. The semiclassical asymptotics of  the
nonstationary Hartree type equation  were also studied
\cite{borisov:Mas1,borisov:Mas2,borisov:Mas3,borisov:Maslov01,
borisov:MSh94,borisov:MSh95,borisov:MSh95a,borisov:MSh98}. This
formalism is based on the WKB-Maslov complex germ method, the idea
of which was first put forward in \cite{borisov:MASLOV-0} and then
comprehensively treated in
\cite{borisov:MASLOV-1,borisov:Mas3a,borisov:BEL-DOB}.

Further developments of the Maslov method
\cite{borisov:27,borisov:28} (see also \cite{borisov:BAGRE} and
references therein) allow one to construct the localized solutions
of the Cauchy problem in the class of trajectory -- concentrated
functions.

Here we apply  the ideas of the complex germ method and the
approach used in \cite{borisov:BTS,borisov:LTS} to construct the
analytic solutions asymptotic in a small parameter $\hbar$,
$\hbar\to 0$, for equation~(\ref{borisov:GPE}). The asymptotic
solutions are constructed in the class of functions localized in
the vicinity of an unclosed surface associated with the phase
curve that describes the evolution of the surface vertex.

The functions of this class have the  one-soliton form of the
one-dimensional nonlinear Schr\"o\-din\-ger equation along the
surface normal direction. A semiclassical linearization of
equation~(\ref{borisov:GPE}) is performed
 accurate to $O(\hbar^{3/2})$,
$\hbar\to 0$. The  asymptotic solutions constructed are
illustrated by examples.

\section{The class of paraboloid-concentrated solitonlike functions}

Let us describe a class of functions $\Psi (\vec x,t,\hbar)$ in
which the asymptotics  for the GPE
 (\ref{borisov:GPE}) are constructed. Following \cite{borisov:BTS,borisov:LTS},
consider equation (\ref{borisov:GPE}) for the  $(1+1)$-dimensional
case with no external field ($U=0$)
\begin{gather}  \label{borisov:NLS} \left( i\hbar
\partial _{t} + \frac{1}{2}\hbar^2 \partial_x^2 + g^{2}|\Psi
( x,t,\hbar)|^{2} \right)\Psi (x,t,\hbar) = 0,
\end{gather}
where $ x\in \mathbb{R}^1$ and $\partial_x=\partial/\partial x$.
Equation (\ref{borisov:NLS}) has a one-soliton solution
\cite{borisov:ZAKHAROV,borisov:ZAKHAROV-1}
\begin{gather}
 \Psi ( x,t,\hbar)= - \frac{2\eta}{g}\,
\frac {1}{{\rm ch} [2\eta (x-x_0- {2\xi}t)/{\hbar }]} \exp
\left[\frac {i}{\hbar}\big(2\xi x-{2} \big(\xi ^2-\eta
^2\big)t+\varphi _0\big)\right], \label{borisov:SOLIT}
\end{gather}
where  $\xi$, $\eta$, $x_0$, and $ \varphi _0$ are the real
parameters.

Earlier \cite{borisov:ST} we  obtained a semiclassical solution of
equation~(\ref{borisov:GPE}) with the leading term
\begin{gather}
\label{borisov:LEADT}  \Psi (\vec
x,t,\hbar)=g^{-1}\sqrt{\langle\nabla\sigma, {\mathcal
H}_{pp}(t)\nabla\sigma\rangle} \exp\left(\frac{i}{\hbar} S(\vec
x,t,\hbar)\right){\rm ch^{-1}}\left(\frac{1}{\hbar}\sigma(\vec
x,t,\hbar)\right).
\end{gather}
It was  assumed that the complex function $\Omega(\vec
x,t,\hbar)=S(\vec x,t,\hbar) +i\sigma(\vec x,t,\hbar)$ regularly
depends on  $\hbar$ and its high-order part, i.e., the function
$\Omega(\vec x,t,0)=$ $\Omega^{(0)}(\vec x,t)=$ $S^{(0)}(\vec x,t)
+i\sigma^{(0)}(\vec x,t)$, is a complex solution of the
Hamilton--Jacobi equation
\begin{gather}
\label{borisov:H-J}  \partial_t \Omega^{(0)}+\mathcal
H\big(\nabla{\Omega^{(0)}},\vec x, t\big)=0.
\end{gather}
 The functions (\ref{borisov:LEADT}) are localized in the  neighborhood of
a $t$-parameter family of surfaces
\begin{gather}
\label{borisov:SURF} \Gamma^t=\big\{\vec x:  \sigma(\vec
x,t,0)=\sigma^{(0)}(\vec x,t)=0\big\}.
\end{gather}
Assume that the rank of  Hesse matrix of the function
$\sigma^{(0)}(\vec x,t)$  is  $n$ at any point $\vec x\in
\Gamma^t$.

The construction of global solutions of equation
(\ref{borisov:H-J}) is beyond the scope of this work. Here we
consider the evolution of the functions localized in the
neighborhood $U_{\vec x_0}$ of the surface point~$\vec
x_0$~(\ref{borisov:SURF}) at the  initial time $t=0$.

Denote by
\[
\gamma=\big\{ \vec x:\vec x\in \mathbb R^n, \vec x=\vec X(t),
t\geqslant0 \big\}
\]
a smooth curve passing through the  point  $\vec x_0$ ($=\vec
X(0)$) in the coordinate space $\vec x\in \mathbb R^n$ such that
$\vec X(t)\in \Gamma^t$. This curve plays the role of the
``classical trajectory'' corresponding to the solution of the
quantum equation (\ref{borisov:GPE}).

In the neighborhood  $\tilde U_{\vec X(t)}\subset \Gamma^t$ of the
point  $\vec X(t)\in \gamma$, the hypersurface $\Gamma^t$ can be
approximated  by a  simpler surface, for example, by the tangent
plane. It is well known that a tangent plane is determined
uniquely by its normal and the point of contact. Let $\vec\pi(t)$
be a normal vector to the surface $\Gamma^t$ at a point $\vec x
=\vec X (t)$.

To construct  asymptotic solutions of equation (\ref{borisov:GPE})
one needs the complex germ \cite{borisov:MASLOV-1}, i.e.\ the
$n$-dimensional complex space associated with the equation under
consideration  and functions defined in the  neighborhood of the
hypersurface (\ref{borisov:SURF}).

It is more precise to use a second order surface
\begin{gather}
\langle\vec \pi(t),\Delta \vec x\rangle +\frac 12 \langle\Delta
\vec x, Q_2(t)\Delta\vec x\rangle=0, \qquad \Delta \vec x=\vec
x-\vec X(t),
 \label{borisov:HYPER}
\end{gather}
instead of the tangent plane. The surface  (\ref{borisov:HYPER})
with a proper choice of the matrix $Q_2(t)$ is a~quadratic
approximation of (\ref{borisov:SURF}) in the neighborhood $\tilde
U_{\vec X(t)}$.

Introduce the   class ${\mathcal N}^t_\hbar$  of complex functions
with the generic element
\begin{gather}
 \Psi (\vec x,t,\hbar)=\Phi
(\theta, \vec x,t,\hbar)=g^{-1}\sqrt{\langle\vec\pi(t), \mathcal
H_{pp}(t)\vec\pi(t)\rangle} \exp\left(\frac{i}{\hbar}
S(\vec x,t,\hbar)\right)\nonumber\\
\phantom{\Psi (\vec x,t,\hbar)=}{}\times {\rm ch}^{-1}(\theta)
\big\{ 1+ \hbar \,{\rm ch}(\theta) u(\theta,\vec x,t,\hbar)+i\hbar
\,{\rm ch}(\theta) v(\theta,\vec x,t,\hbar)
\big\}.\label{borisov:GENEL}
\end{gather}
The functions (\ref{borisov:GENEL})  are localized in the
neighborhood $ U_{\vec X(t)}$ of the point $\vec x=\vec X(t)$.
This point is the ``vertex'' of the surface (\ref{borisov:HYPER}).
In equation (\ref{borisov:GENEL}),  $\theta=\sigma(\vec
x,t,\hbar)/{\hbar}$ is a  ``fast''  variable; the real functions
$u=u(\theta,\vec x,t,\hbar)$ and $v=v(\theta,\vec x,t,\hbar)$ are
regular in their arguments and are bounded in~$\theta$, and
\begin{gather}
\sigma(\vec x,t,\hbar)=\langle\vec \pi(t),\Delta \vec x\rangle
+\frac 12 \langle\Delta \vec x, Q_2(t)\Delta \vec x\rangle +
\hbar\sigma^{1}(\vec x,t,\hbar),
 \label{borisov:SIGMA}
\\
S(\vec x,t,\hbar)=S^{0}(t)+\langle\vec P(t),\Delta\vec x\rangle
+\frac{1}{2}\langle\Delta\vec x, Q_1(t)\,\Delta\vec x\rangle+\hbar
S^{1}(\vec x,t,\hbar). \label{borisov:S}
\end{gather}
In equations (\ref{borisov:SIGMA}) and (\ref{borisov:S}), the
real function $S^0(t)$, the real $n$-dimensional vector functions
$\vec\pi(t)$, $\vec P(t)$, and $\vec X(t)$, the real $(n\times
n)$-matrix functions  $Q_1(t)$ and $Q_2(t)$, the real functions
$\sigma^1(\vec x,t,\hbar)=\tilde\sigma^1(\vec\xi,t,\hbar)$ and
$S^1(\vec x,t,\hbar)=\tilde S^1(\vec\xi,t,\hbar)$,
$\vec\xi={\vec\Delta x}/{\sqrt\hbar}$ are functional parameters of
the class ${\mathcal N}^t_\hbar$; $ \Delta \vec x=\vec x-\vec
X(t)$.

Note that  in constructing   the  class ${\mathcal N}^t_\hbar$, a
specific  basis,  ``orthogonal''  in some way to the vector
$\vec\pi(t)$, is used\footnote{Such a situation is typical of the
complex germ method \cite{borisov:MASLOV-1,borisov:BEL-DOB} where
the solution of spectral problem is considered for the functions
localized on incomplete Lagrange manifolds with the use of a germ
basis  which is ``tangent'' to the manifold.}. As a result, the
classical trajectory depends  on the complex germ as well.

The functions (\ref{borisov:GENEL}) are not normalizable in the
space $L_2({\mathbb R}^n,d\vec x)$ since they are localized on an
unclosed  surface, equation~(\ref{borisov:HYPER}). Therefore, the
surface $\Gamma^t$ (\ref{borisov:SURF}) (compact in some physical
applications) is substituted by its quadratic approximation
(\ref{borisov:HYPER}) in the neighborhood $ U_{\vec X(t)}$. As we
are interested in finding functions localized in $ U_{\vec X(t)}$,
we come to the  problem of small perturbations on the
``background'' of the hypersurface (\ref{borisov:HYPER}). To
eliminate the ``background'', consider the function
\begin{gather}
  \tilde\Psi (\vec
x,t,\hbar)=\Phi (\theta, \vec x,t,\hbar)-\Phi (\theta_0, \vec
x,t,\hbar)\thickapprox g^{-1}\frac d{d\theta}\sqrt{
{\langle\vec\pi, {\mathcal
H}_{pp}(t)\vec\pi\rangle}}\,{\rm ch}^{-1}(\theta) \nonumber\\
\phantom{\tilde\Psi (\vec x,t,\hbar)=}{} \times
\exp\left(\frac{i}{\hbar} S(\vec x,t,\hbar)\right) \left\{ 1+\hbar
\,{\rm ch}(\theta) u(\theta,\vec x,t,\hbar)+  i\hbar \,{\rm
ch}(\theta) v(\theta,\vec x,t,\hbar)
\right\}\Bigg|_{\theta=\theta_0}\!\!\theta_1,\!\!\!\label{borisov:BACKGR} \\
\theta=\theta_0+\theta_1,\qquad\theta_0=\frac1\hbar\langle\vec
\pi(t),\Delta \vec x\rangle+ \frac 1{2\hbar} \langle\Delta \vec x,
Q_2(t)\Delta \vec x\rangle, \qquad \theta_1=  \tilde
\sigma^{1}(\vec\xi,t,\hbar).\nonumber
\end{gather}

\begin{figure}[t]
\centerline{\includegraphics[width=2in]{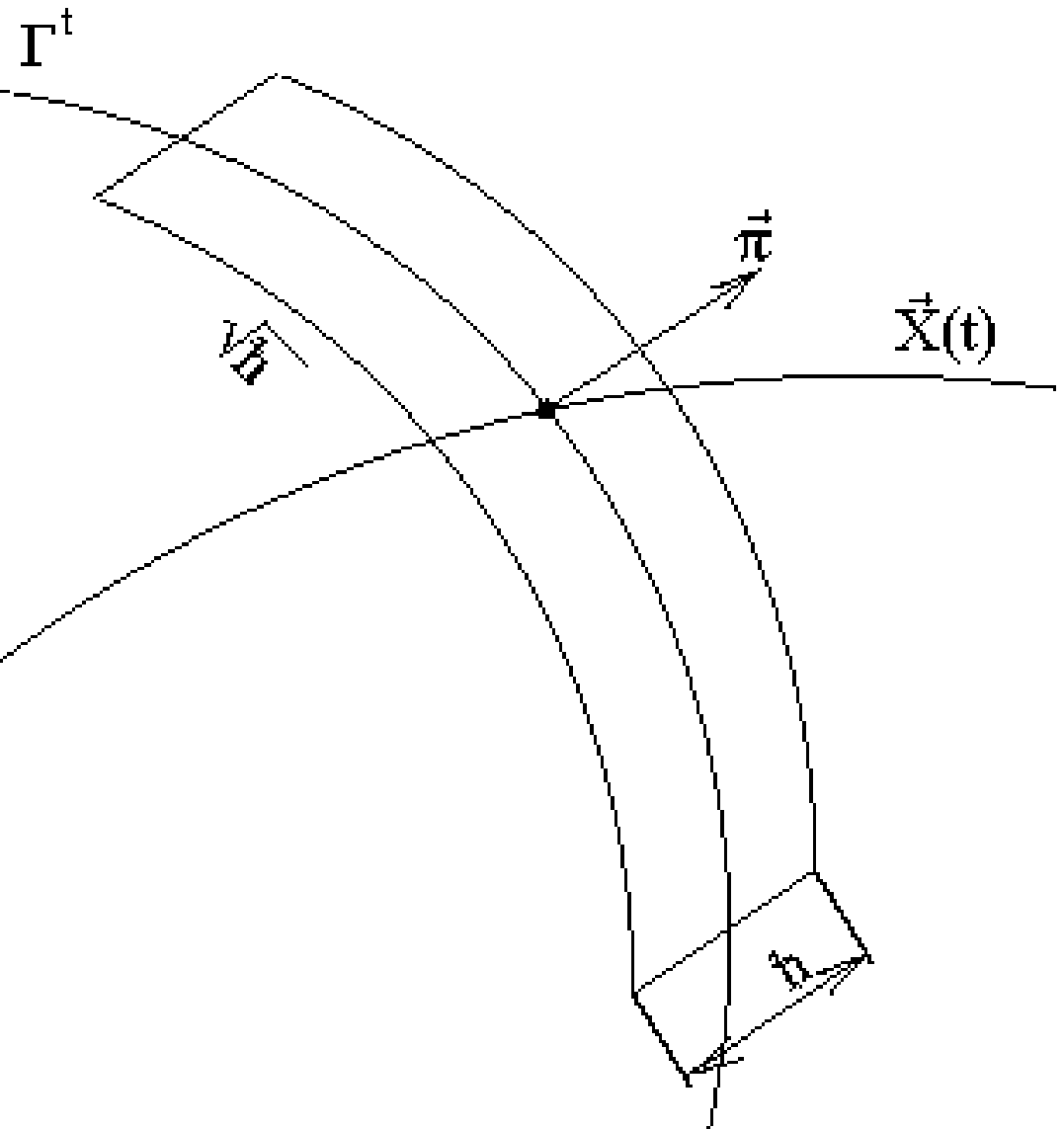}}

\vspace{-3mm}

\caption{} \label{borisov:Fig1}
\end{figure}

The functions  $\tilde\Psi (\vec x,t,\hbar)$,   equation
(\ref{borisov:BACKGR}), are normalizable in  $L_2({\mathbb
R}^n,d\vec x)$. Estimating the  solutions of
equation~(\ref{borisov:GPE}), we use  the norm  of the functions
$\tilde\Psi (\vec x,t,\hbar)$. Taking into account this
normalization condition, let us assume that the functions $\tilde
\sigma^1(\vec\xi,t,\hbar)$ and $\tilde S^1(\vec\xi,t,\hbar)$
belong to the Schwartz space  ${\mathcal S}({\mathbb
R}^{n-1}_\xi,d\mu)$, where $d\mu$ is the measure on the
hypersurface, determined by equation~(\ref{borisov:HYPER}).

The class ${\mathcal N}^t_\hbar$  of the form
(\ref{borisov:GENEL}) with $n=1$ includes the exact one-soliton
solution (\ref{borisov:SOLIT}) of   one-dimensional nonlinear
Schr\"odinger equation  (\ref{borisov:NLS}) with a special choice
of the functional parameters of the class. If the rank of  matrix
$Q_2(t)$ is $n$, then in  one-dimensional case
equation~(\ref{borisov:GENEL}) can not be transformed  to
(\ref{borisov:SOLIT}) since the argument $\theta$  of the
hyperbolic cosine contains the terms quadratic in~$x$.
 The sufficient conditions for such a transformation are
\begin{gather}
 \label{borisov:RANK} {\rm rank}\, (Q_2(t))=n-1, \qquad
{\rm rank}\, (Q_2(t), \vec\pi(t))=n,
\end{gather}
where  $(Q_2(t), \vec\pi(t))$ is the augmented matrix of  order
$n\times(n+1)$. Note that under conditions~(\ref{borisov:RANK}),
the surface (\ref{borisov:HYPER}) has the paraboloidal shape.

Let us expand the operator $\widehat{L}$ in
equation~(\ref{borisov:GPE}) in the neighborhood $ U_{\vec X(t)}$
under the conditions
\begin{gather}\label{borisov:ESTIM}
\Delta x_k=\hat O\big(\hbar^{1/2}\big), \qquad \Delta\hat
p_k\Big|_{\theta=\mathop{\rm const}}=\hat O\big(\hbar^{1/2}\big),
\qquad\langle\vec\pi(t),\Delta\vec x\rangle=\hat O(\hbar).
\end{gather}
The  leading term of the asymptotic in the  class of functions
(\ref{borisov:GENEL}) under  conditions (\ref{borisov:ESTIM}) has
the form  (\ref{borisov:LEADT}).

It is determined by the phase curve $z=Z(t,\hbar)=(\vec
P(t,\hbar), \vec X(t,\hbar))$,  vector $\vec\pi=\vec\pi(t)$, and
functions $\sigma(\vec x,t,\hbar)$ and $S(\vec x,t,\hbar)$ of the
form  (\ref{borisov:SIGMA}) and (\ref{borisov:S}), respectively.
We call these functions the {\it  paraboloid-concentrated
solitonlike functions}.

\section{The linear associated Schr\"odinger equation}

The asymptotic leading term  is obtained when the asymptotic
solution of the equation (\ref{borisov:GPE}) is constructed
accurate to $O(\hbar^2)$ \cite{borisov:ST}.

Let us substitute (\ref{borisov:GENEL}) in (\ref{borisov:GPE}),
take into account the  terms of  order $O(\hbar^2)$, separate the
equations with respect to the ``fast'' variable, and find the
solutions of these equations that  decrease as
$\theta\to\pm\infty$. We than obtain for the function $\Omega
=S+i\sigma$
\begin{gather}
  \Omega_t+{\mathcal H}(\nabla \Omega,\vec x,t)-
\frac{i\hbar}2\bigg[
 \mathop{\rm Sp}[{\mathcal H}_{pp}(t)\Omega_{xx}+{\mathcal H}_{px}(\nabla \Omega, \vec
x,t)]\nonumber\\
\phantom{\Omega_t}{}+ {\langle\vec\pi, {\mathcal
H}_{pp}(t)\vec\pi\rangle}^{-1}\frac d{dt}\langle\vec\pi, {\mathcal
H}_{pp}(t)\vec\pi\rangle \bigg]=0.\label{borisov:LEQ-1}
\end{gather}
With the substitution $\Omega=-i\hbar\ln\left[ {\psi(\vec
x,t,\hbar)}{ {\langle\vec\pi, {\mathcal
H}_{pp}(t)\vec\pi\rangle}^{-1/2}}\right] $,
equation~(\ref{borisov:LEQ-1}) yields
\[
 \big[ -i\hbar
\partial _{t} + {\mathcal H}(\hat {\vec p},\vec x,t) \big]
\psi (\vec x,t,\hbar) = O\big(\hbar^{3/2}\big),
\]
where the operator ${\mathcal H}(\hat {\vec p},\vec x,t)$ has  the
form of (\ref{borisov:GPE1}),
\begin{gather}
\psi(\vec x,t,\hbar)=\sqrt{\langle\vec\pi, \mathcal
H_{pp}(t)\vec\pi\rangle}
\exp\left[{\frac{i}{\hbar}\Omega(\vec
x,t,\hbar)}\right]=
\exp\left[\frac{i}{\hbar}S(\vec x,t,\hbar)\right]\varphi(\vec x,t, \hbar),\label{borisov:LEQ-3}\\
 \varphi(\vec x,t, \hbar)=\sqrt{\langle\vec\pi, {\mathcal
H}_{pp}(t)\vec\pi\rangle} 
\exp\left[{-\frac{1}{\hbar}\sigma(\vec
x,t,\hbar)}\right].\nonumber
\end{gather}

We call the equation
\begin{gather} \label{borisov:LASE}
 \big[ -i\hbar
\partial _{t} + {\mathcal H}(\hat {\vec p},\vec x,t) \big]
\psi (\vec x,t,\hbar) = 0
\end{gather}
the {\em linear associated Schr\"odinger equation} corresponding
to the GPE~(\ref{borisov:GPE}).

Therefore, the  leading term of the asymptotic solution of the
nonlinear equation (\ref{borisov:GPE}) is constructed by using a
solution of the linear equation (\ref{borisov:LASE}) with
conditions (\ref{borisov:LEADT}) and (\ref{borisov:LEQ-3}).

Note that  for the function (\ref{borisov:LEADT}) to satisfy
(\ref{borisov:RANK})  the solution (\ref{borisov:LEQ-3}) must be
chosen in a specific way. Denote the class of these  functions by
${\mathcal S}_\hbar^t$.

Let us seek  the solutions to equation (\ref{borisov:LASE}) that
possess this property in the  class of functions
\begin{gather}
\psi (\vec x,t,\hbar)=\exp\left[-\frac{1}{\hbar}\langle\vec\pi,
\Delta\vec x\rangle\right]\phi (\vec
x,t,\hbar).\label{borisov:FUNC}
\end{gather}

Substitution of  (\ref{borisov:FUNC}) in   (\ref{borisov:LASE})
yields
\begin{gather}
 \big[ -i\hbar
\partial _{t} +i\langle\dot{\vec\pi},
\Delta\vec x\rangle- i\langle{\vec\pi}, \dot{\vec
x}\rangle+{\mathcal H}\big(\vec p(t)+i\vec \pi+\Delta\hat {\vec
p},\vec x,t\big) \big] \phi (\vec x,t,\hbar) =
O\big(\hbar^{3/2}\big).\label{borisov:LEQ-4}
\end{gather}
Let us expand the operator of equation (\ref{borisov:LEQ-4}) in a
power series in $\Delta\vec x$ accurate to
$O\big(\hbar^{3/2}\big)$ in terms of the accuracy of
(\ref{borisov:ESTIM}). We then  have
\begin{gather}
  \Big[ -i\hbar
\partial _{t} -i\big[\langle{\vec\pi}, (\dot{\vec
x}-{\mathcal H}_{\vec
p}(t))\rangle-\langle(\dot{\vec\pi}+{\mathcal H}_{xp}(t)\vec\pi),
\Delta\vec x\rangle-\langle{\vec\pi, {\mathcal H}_{pp}(t)
\Delta\hat{\vec p}}\rangle\big]\nonumber\\
\qquad {}+{\mathcal H}(t)- \frac 12\langle{\vec\pi, {\mathcal
H}_{pp}(t) \vec\pi}\rangle +\langle{{\mathcal H}_{\vec x}(t),
\Delta\vec x}\rangle + \langle{{\mathcal H}_{\vec p}(t),
\Delta\hat{\vec
p}}\rangle \nonumber\\
\qquad{}+\frac12\Big[\langle{\Delta \vec x, {\mathcal
H}_{xx}(t)\Delta\vec x}\rangle+\langle{\Delta \vec x, {\mathcal
H}_{xp}(t) \Delta\hat{\vec p}}\rangle +\langle{\Delta\hat{\vec p},
{\mathcal H}_{px}(t)\Delta \vec
x}\rangle \nonumber\\
\qquad {} +\langle{\Delta\hat{\vec p}, {\mathcal
H}_{pp}(t)\Delta\hat{\vec p}}\rangle\big]\Big] \phi  +
O\big(\hbar^{3/2}\big)=\big[ -i\hbar
\partial _{t} +\hat{\mathcal H}_0(t)\big] \phi  + O\big(\hbar^{3/2}\big)=0.
\label{borisov:LEQ-5}
\end{gather}
Therefore, for the solution (\ref{borisov:FUNC}) in the
approximation under consideration, the linear associated
Schr\"odinger equation (\ref{borisov:LASE}) takes the form
\begin{gather}
\big[ -i\hbar
\partial _{t} +\hat{\mathcal H}_0(t)\big] \phi=0.
\label{borisov:LASE-1}
\end{gather}
Here the  operator $\hat{\mathcal H}_0(t)$ (quadratic with respect
to the operators $\Delta{\vec x}$ and $\Delta\hat{\vec p}\,$) is
obtained from~(\ref{borisov:LEQ-5}).

Thus, the  construction  of  asymptotic leading term
(\ref{borisov:LEADT}) in the class of functions
(\ref{borisov:FUNC}) for the nonlinear equation
(\ref{borisov:GPE})
 by  solving   equation~(\ref{borisov:LASE-1}) is complete.

\section[Solutions of the Gross-Pitaevskii equation]{Solutions of the Gross--Pitaevskii equation}
\label{borisov:SecGPE}

The solutions of equation (\ref{borisov:LASE-1}) are well-known
(see, e.g., \cite{borisov:MANKO,borisov:75,borisov:76}). In
particular, the evolution operator $\hat U_{{\mathcal H}_0}(t,s)$
is found in explicit form, whose action on the function $\psi
(\vec x,\hbar)$ referred to a~time $t=s$ is given by the relation
\begin{gather}
\psi (\vec x,t,\hbar)=\hat U_{{\mathcal H}_0}(t,s)\psi(\vec
x,\hbar)= \int_{{\mathbb R}^n}G_{{\mathcal H}_0}(\vec x,\vec
y,t,s)\psi (\vec y,\hbar)d\vec y. \label{borisov:EVOL-1}
\end{gather}
Here $G_{{\mathcal H}_0}(\vec x,\vec y,t,s)$ is the Green function
of the linear associated Schr\"odinger equation
(\ref{borisov:LASE-1}), which is determined by the following
conditions:
\begin{eqnarray}
\label{borisov:GREEN-1} \big\{-i\hbar\partial_t + \hat{\mathcal
H}_0(t) \big\} G_{{\mathcal H}_0}(\vec x,\vec y,t,s)=0, \qquad
\lim_{t\to s}G_{{\mathcal H}_0}(\vec x,\vec y,t,s)=\delta(\vec
x-\vec y).
\end{eqnarray}

To construct a subclass of the class ${\mathcal N}^t_\hbar$
(\ref{borisov:GENEL}) of solutions of
equation~(\ref{borisov:GPE}), let us find a particular solution in
the form
\begin{gather}
\phi (\vec x,t)=N_\phi\exp\left\{\frac i\hbar\left[S^0(t)+\hbar
S^1(t)+i\hbar\sigma^1(t)+ \langle\vec P(t),\Delta\vec
x\rangle+\frac 12\langle\Delta\vec x,Q(t)\Delta\vec x\rangle
\right]\right\},\label{borisov:VACUUM}
\end{gather}
where the functions $S^0(t)$, $S^1(t)$, and $\sigma^1(t)$, the
$n$-dimensional vectors  $\vec P(t)$, $\vec X(t)$, and the complex
$(n\times n)$-matrix $Q(t)$=$Q_1(t)+iQ_2(t)$ are to be determined;
$N_\phi$ is a constant.

Substituting (\ref{borisov:VACUUM}) in (\ref{borisov:LEQ-5}) and
setting $\dot{\vec X}={\mathcal H}_{\vec p}$,
$Q(t)=Q_1(t)+iQ_2(t)$, we obtain the determining system of
equations for $\vec P(t)$, $\vec \pi(t)$, $\vec X(t)$, $Q(t)$
\begin{gather}
 \dot{\vec P}=-{\mathcal H}_{\vec x}(\vec P, \vec
X,t)+
Q_2{\mathcal H}_{pp} (t)\vec \pi, \label{borisov:HAMILT}\\
 \dot{\vec X}={\mathcal H}_{\vec p}(\vec P, \vec X,t), \nonumber\\
 \dot{\vec \pi}=-\big[{\mathcal H}_{xp}
\big(\vec P, \vec X,t\big)+Q_1(t){\mathcal H}_{pp}
(t)\big]\vec\pi, \nonumber\\
 \dot Q+{\mathcal
H}_{xx}(t)+ {\mathcal H}_{xp}(t)Q+ Q{\mathcal
H}_{px}(t)+ Q{\mathcal
H}_{pp}(t)Q=0.\label{borisov:SYSVARq}
\end{gather}

Consider the following system of equations
\begin{gather}
 \dot{\vec Z}= {\mathcal H}_{px}(\vec P, \vec
X,t)\vec Z + \mathcal H_{pp}(t)\vec W, \qquad \dot{\vec W}=
-{\mathcal H}_{xx}(\vec P, \vec X,t)\vec Z - {\mathcal
H}_{xp}(\vec P, \vec X,t)\vec W. \label{borisov:SYSVAR}
\end{gather}
 Equations~(\ref{borisov:SYSVAR}) are called a {\it  system in variations}
 in  vector form \cite{borisov:BAGRE}.

 In general, it has $n$ complex linear independent
solutions, which can be written in the form of $2n$-dimensional
vector columns
\begin{gather}
a_j(t)=\big(\vec W_j(t), \vec Z_j(t)\big)^\intercal, \qquad
j=\overline{1,n}. \label{borisov:VECCOLUMN}
\end{gather}
For equation (\ref{borisov:LASE-1}) the vectors
(\ref{borisov:VECCOLUMN}) set the symmetry operators which are
linear in coordinates and momenta
\begin{gather}
\hat a_j(t)=N_j^a\big[\langle\vec Z_j(t),\Delta\hat{\vec
p}\,\rangle- \langle\vec W_j(t),\Delta\vec x\rangle\big], \qquad N_j^a={\rm const},
\label{borisov:SYMMOPER}
\end{gather}
 since the conditions
\begin{gather}
-i\hbar\frac{\partial\hat a_j(t)}{\partial t}+[\hat{\mathcal
H}_0(t),\hat a_j(t)]=0\label{borisov:SYMMOPER-1}
\end{gather}
are satisfied because of the validity  of equation
(\ref{borisov:SYSVAR});
 $[\hat{\mathcal H}_0(t),\hat a_j(t)]= \hat{\mathcal H}_0(t)\hat
a_j(t)-\hat a_j(t)\hat{\mathcal H}_0(t)$ is  the commutator of the
linear operators.

Let $B(t)$ and $C(t)$ denote the $(n\times n)$-matrices  whose
columns are constructed of the vectors of the solutions of system
(\ref{borisov:SYSVAR}):
\[
B(t)=(\vec W_1(t),\dots,\vec W_n(t)),\qquad  C(t)=(\vec
Z_1(t),\dots,\vec Z_n(t)).
\]
Then $Q(t)=B(t)C^{-1}(t)$ and  $ \vec W(t)=$ $Q(t)\vec Z(t)$, and
the system in variations (\ref{borisov:SYSVAR}) is equivalent to
(\ref{borisov:SYSVARq}) (see \cite{borisov:BAGRE}).

The normalization factors $N_j^a$ are chosen as follows. The
matrices $B(t)$ and $C(t)$ satisfy the condition
\[
D_0(t)=\frac1{2i}[C^+(t)B(t)-B^+(t)C(t)]= D_0(0)=D_0.
\]
Here $C^+$ is  Hermitian adjoint  to $C$. In addition, if the
matrix $Q(0)=B(0)C^{-1}(0)$ is  symmetric at the initial time
$t=0$, $Q(0)=Q(0)^\intercal$, then $Q(t)=Q(t)^\intercal$, $t>0$.

Let
\[
D_0= \mbox{diag}(d_1,d_2,\dots , d_{n-1},0),\qquad d_j>0, \qquad
N_j^a=(2\hbar d_j)^{-1/2}, \qquad N_{n}^a=1.
\]
Then the operators (\ref{borisov:SYMMOPER}) satisfy the ordinary
commutation relations
\[
[\hat a_j(t),\hat a_k^+(t)]=\delta_{jk},\qquad
j,k=\overline{1,n-1},
\]
and all the other commutators are equal to zero.

Let us denote the function $\psi(\vec x,t,\hbar)$ in
(\ref{borisov:FUNC}) as
\begin{gather}
\psi_0^{(0)}(\vec x,t,\hbar)= N_0\exp\left[-\frac
1\hbar\langle\vec\pi(t), \Delta\vec x\rangle\right]|0,t\rangle,
\label{borisov:VACUUM1}
\end{gather}
where the function $\phi(\vec x,t,\hbar)$ of the form
(\ref{borisov:VACUUM}) is denoted as $|0,t\rangle$. The
normalization factor $N_0$ is determined from the normalization
condition for the  function $\Psi(\vec x,t,\hbar)$ that is a
solution of the nonlinear equation (\ref{borisov:GPE}).

Let us substitute  (\ref{borisov:VACUUM1}) in
(\ref{borisov:LASE-1}); then
\begin{gather*}
{\psi_0^{(0)}}(\vec x,t,\hbar)={N_0}\sqrt{\frac{\det C(0)}{\det
C(t)}}  \exp\left[-\frac
1\hbar\langle\vec\pi(t), \Delta\vec x\rangle\right]\\
\phantom{{\psi_0^{(0)}}(\vec x,t,\hbar)=} {}\times\exp\Bigg\{\frac
i\hbar\left[\int_0^t \left[\langle\vec P(t),\dot{\vec
X}(t)\rangle- {\mathcal H}(t)+\frac 12 \langle\vec\pi(t),
{\mathcal
H}_{pp}(t)\vec\pi(t)\rangle\right]dt\right.\\
\phantom{{\psi_0^{(0)}}(\vec x,t,\hbar)=} {} \left. +\langle\vec
P(t),\Delta\vec x\rangle + \frac 12\langle\Delta\vec
x,Q(t)\Delta\vec x\rangle \right]\Bigg\}.
\end{gather*}
Here  $(\vec P(t),\vec X(t))$ is a solution of the first pair of
equations  (\ref{borisov:HAMILT}), the $(n\times n)$-matrix $C(t)$
is composed of the columns $\vec Z_j(t)$. Note in addition that
$\langle\Delta\vec x, {\rm Im}Q(t)\Delta\vec x\rangle\ge 0$ and
\begin{gather}
\hat a_j(t) |0,t\rangle=0, \qquad j,k =\overline{1,n-1}.
\label{borisov:VAC-5}
\end{gather}
We have for the function $\psi_0^{(0)}(\vec x,t,\hbar)$
\begin{gather}
\hat{ \tilde a}_n(t) \psi_0^{(0)}(\vec x,t,\hbar)=0, \label{borisov:VACUUM3}\\
\hat{ \tilde a}_n(t)=\exp\left[\frac 1\hbar\langle\vec\pi(t),
\Delta\vec x\rangle\right] \hat{a}_n(t)\exp\left[-\frac
1\hbar\langle\vec\pi(t), \Delta\vec x\rangle\right]
=\hat{a}_n(t)-i.\nonumber
\end{gather}
Let  $\Psi_0(\vec x,t,\hbar)$ denote the function $\Psi(\vec
x,t,\hbar)$, Eq.  (\ref{borisov:LEADT}), corresponding to the
function $\psi_0^{(0)}(\vec x,t,\hbar)$:
\begin{gather}
\Psi_0 (\vec x,t,\hbar)=\sqrt{\frac {\langle\vec\pi, {\mathcal
H}_{pp}(t)\vec\pi\rangle}{g^{2}}}\frac{\exp\left(\frac{i}{\hbar}
S_0(\vec x,t,\hbar)\right)}{{\rm ch}\left(\frac{1}{\hbar}
\sigma_0(\vec x,t,\hbar)\right)}.\label{borisov:VACUUM4}
\end{gather}
In accordance with (\ref{borisov:LEQ-3}), we have
\begin{gather}
S_0(\vec x,t,\hbar)=\int_0^t \left[\langle\vec P(t),\dot{\vec
X}(t)\rangle- {\mathcal H}(t)+\frac12 \langle\vec\pi(t), {\mathcal
H}_{pp}(t)\vec\pi(t)\rangle\right]dt\nonumber\\
\phantom{S_0(\vec x,t,\hbar)=}{}+\langle\vec P(t),\Delta\vec
x\rangle+\frac12\langle\Delta\vec x,Q_1(t)\Delta\vec
x\rangle+\hbar\,{\rm Im}\left[\ln\sqrt{\frac{\det C(0)}{\det
C(t)}}\,\right] +\hbar\,{\rm Im}\, N_0,\label{borisov:SOLUT-3a}\\
\sigma_0(\vec x,t,\hbar)=\langle\vec\pi(t), \Delta\vec x\rangle
+\frac12\langle\Delta\vec x,Q_2(t)\Delta\vec x\rangle
+\hbar\ln\sqrt{\langle\vec\pi(t),{\mathcal
H}_{pp}(t) \vec\pi(t)\rangle} \nonumber\\
\phantom{\sigma_0(\vec x,t,\hbar)=}{}-\hbar\,{\rm Re}\left[\ln
\sqrt{\frac{\det C(0)}{\det C(t)}}\right] -\hbar\,{\rm Re}\, N_0.
\label{borisov:SOLUT-4}
\end{gather}
Equations (\ref{borisov:VACUUM4})--(\ref{borisov:SOLUT-4})
determine the leading term of the asymptotic solution of
(\ref{borisov:GPE}) that corresponds to the solution
$\psi_0^{(0)}(\vec x,t,\hbar)$
of the form  (\ref{borisov:LEQ-3}) for the linear associated
Schr\"odinger equation~(\ref{borisov:LASE}).

Let us introduce a set of functions $|\nu,t\rangle$ with the help
of  ``creation operators'' $\hat a_j^+(t)$, $j=\overline{1,n-1}$,
acting on the function $|0,t\rangle$:
\begin{gather}
 \psi_\nu^{(0)}(\vec x,t,\hbar)=N_{\nu}\exp\left[-
\frac1\hbar\langle\vec\pi(t), \Delta\vec
x\rangle\right] |\nu,t\rangle\nonumber\\
\phantom{\psi_\nu^{(0)}(\vec x,t,\hbar)}{}=N_{\nu}\exp\left[-\frac
1\hbar\langle\vec\pi(t), \Delta\vec
x\rangle\right]\frac {1}{\sqrt{\nu!}}[\hat{\vec a}{}^+(t)]^\nu |0,t\rangle\nonumber\\
\phantom{\psi_\nu^{(0)}(\vec x,t,\hbar)}{}=N_{\nu}\exp\left[-\frac
1\hbar\langle\vec\pi(t), \Delta\vec
x\rangle\right]\prod_{j=1}^{n-1}\frac {1}{\sqrt{\nu_j!}}[\hat a_j^+(t)]^{\nu_j}|0,t\rangle\nonumber\\
\phantom{\psi_\nu^{(0)}(\vec x,t,\hbar)}{}={N_{\nu}}{N_0}^{-1}
H_\nu(\vec x,t,\hbar) \psi_0^{(0)}(\vec x,t,\hbar).
\label{borisov:SOLUT-5}
\end{gather}
Here  $\nu=(\nu_1,\nu_2,\dots,\nu_{n-1})\in{\mathbb Z}_+^{n-1}$ is
a mutiindex, $ \hat{\vec a}{}^+(t)=$ $(\hat a^+_1(t),\dots,\hat
a^+_{n-1}(t))$, and $H_\nu(\vec x,t,\hbar)$ are many-dimensional
Hermite polynomials. The normalization factors $N_\nu$ are
determined from the normalization conditions similar to those for
$N_0$ in (\ref{borisov:VACUUM1}).

Using expressions similar to
(\ref{borisov:VACUUM4})--(\ref{borisov:SOLUT-4}), we obtain the
function
\begin{gather}
 \label{borisov:SOLUT-6} \Psi_\nu (\vec
x,t,\hbar)=\sqrt{\frac  {\langle\vec\pi, {\mathcal
H}_{pp}(t)\vec\pi\rangle}{g^{2}}} \frac{\exp\left(\frac{i}{\hbar}
S_{\nu}(\vec x,t,\hbar)\right)}{{\rm ch}\left(\frac{1}{\hbar}
\sigma_{\nu}(\vec x,t,\hbar)\right)},\\
S_\nu(\vec x,t,\hbar)=S_0(\vec x,t,\hbar)+\hbar\,{\rm
Im}\,\bigl[\ln H_\nu(\vec
x,t,\hbar)\bigr]+\hbar\,{\rm Im}\, (N_\nu-N_0),\label{borisov:SOLUT-7}\\
\sigma_\nu(\vec x,t,\hbar)=\sigma_0(\vec x,t,\hbar)-\hbar\,{\rm
Re}\,\bigl[\ln H_\nu(\vec x,t,\hbar)\bigr]-\hbar\, {\rm Re}\,
(N_\nu-N_0)\label{borisov:SOLUT-8}
\end{gather}
for the function (\ref{borisov:LEADT}) corresponding to
$\psi_\nu^{(0)}(\vec x,t,\hbar)$.

Let us define  class $\mathcal K^t_\hbar$ of solutions for the
linear associated Schr\"odinger equation (\ref{borisov:LASE}) as
a~linear envelope of the functions $\psi_\nu^{(0)}(\vec
x,t,\hbar)$. The generic element of  class $\mathcal K^t_\hbar$ is
given by
\begin{gather*}
 \psi_{\mathcal K}(\vec x, t,\hbar)=\sum_{|\nu|=0}^\infty c_\nu\psi_\nu^{(0)}(\vec x,t,\hbar)=
\exp\left(-\frac {1}{\hbar}\langle\vec\pi(t),\vec\Delta x \rangle
\right)
\phi_{\mathcal K}(\vec x, t, \hbar)\nonumber\\
\phantom{\psi_{\mathcal K}(\vec x, t,\hbar)}{}=\exp\left(-\frac
{1}{\hbar}\langle\vec\pi(t),\vec\Delta x \rangle \right)
\sum_{|\nu|=0}^\infty c_\nu |\nu,t\rangle.\nonumber
\end{gather*}
The condition
\begin{gather}
\hat{\tilde a}_n(t)\psi_{\mathcal K}(\vec x, t, \hbar)=
\hat{a}_n(t)\phi_{\mathcal K}(\vec x, t, \hbar)=0
\label{borisov:VACUUM5}
\end{gather}
is valid for the functions of the class $\mathcal K^t_\hbar$. This
follows immediately from (\ref{borisov:VAC-5}),
(\ref{borisov:VACUUM3}), and (\ref{borisov:SOLUT-5}).

Let $\phi_{\mathcal K}(\vec x, s, \hbar) $ be a function of the
class $\mathcal K^s_\hbar$ referred to a start time $s$, and $\hat
U_{\mathcal H_0}(t,s)$ is the evolution operator given by
(\ref{borisov:EVOL-1}) and (\ref{borisov:GREEN-1}).

Then the function
\[
\psi_{\mathcal K}(\vec x, t, \hbar)=\exp\left(-\frac
{1}{\hbar}\langle\vec\pi(t),\vec\Delta x \rangle \right) \hat
U_{\mathcal H_0}(t,s)\phi_{\mathcal K}(\vec x, s, \hbar)
\]
also belongs  to the class $\mathcal K^t_\hbar$, since the
function
\begin{gather}
\label{borisov:EVOL-7A} \phi_{\mathcal K}(\vec x, t, \hbar)=\hat
U_{\mathcal H_0}(t,s)\phi_{\mathcal K}(\vec x, s, \hbar)
\end{gather}
is represented as
\[
\phi_{\mathcal K}(\vec x, s, \hbar)=\sum_{|\nu|=0}^\infty c_\nu
|\nu,t\rangle.
\]
The above relation follows from the uniqueness of  solution of the
Cauchy problem for equation~(\ref{borisov:LASE-1}). On the other
hand, we have for the symmetry operator $\hat a_n(t)$ of equation
(\ref{borisov:LASE-1}) the relation
\[
\hat a_n(t)=\hat U_{\mathcal H_0}(t,s) \hat a_n(s)\hat
U^{-1}_{\mathcal H_0}(t,s)
\]
following from (\ref{borisov:EVOL-1}) and
(\ref{borisov:SYMMOPER-1}). If the function $\phi_{\mathcal
K}(\vec x, s, \hbar)$ belongs to the class $\mathcal K^s_\hbar$ at
the initial time $s$ and  satisfies the condition
\begin{gather}
\hat{a}_n(s)\phi_{\mathcal K}(\vec x, s,
\hbar)=0,\label{borisov:EVOL-5A}
\end{gather}
then  condition (\ref{borisov:VACUUM5}) is also satisfied for the
function (\ref{borisov:EVOL-7A}) at a time $t>s$, i.e., we have
\[
\hat{a}_n(t)\phi_{\mathcal K}(\vec x, t, \hbar)=0.
\]
Therefore, the function $\phi_{\mathcal K}(\vec x, t, \hbar)$
belongs to the class  $\mathcal K^t_\hbar$.

The evolution operator $\hat U_{\mathcal H_0}(t,s)$ of  equation
(\ref{borisov:LASE-1}) induces the evolution operator $\hat U_{\rm
tr}(t,s,\cdot)$ for the nonlinear equation (\ref{borisov:GPE}) in
the class  ${\mathcal S}^t_\hbar$.

Let a function $\psi(\vec x,\hbar)$ be referred to an initial
time~$s$.
%
%
According to (\ref{borisov:LEQ-3}), the function $\Psi_s (\vec
x,\hbar)$ corresponds to $\psi(\vec x,\hbar)$. Then the function
$\Psi (\vec x,t,\hbar)$, determined by (\ref{borisov:EVOL-1}),
corresponds to $\psi (\vec x,t,\hbar)$. This correspondence can be
written as a result of the action of the evolution operator $\hat
U_{\rm tr}(t,s, \cdot)$ on the initial function $\Psi_s (\vec
x,\hbar)$,
\begin{gather}
 \Psi (\vec x,t,\hbar)= \hat U_{\rm tr}(t,s, \Psi_s) (\vec x,t,\hbar).
 \label{borisov:EVOL-NL1}
\end{gather}
Using  the evolution operators $\hat U_{\mathcal H_0}(t,s, \cdot)$
and $\hat U_{\hat L}(t,s, \cdot)$ of the forms
(\ref{borisov:EVOL-1}) and (\ref{borisov:EVOL-NL1}) respectively,
we can define the symmetry operators for the nonlinear equation
(\ref{borisov:GPE}) in the class of functions~${\mathcal
S}^t_\hbar$. To that end, let us take the function $\varphi(\vec
x,\hbar)$ from the class ${\mathcal K}^s_\hbar$  which  satisfies
 condition (\ref{borisov:EVOL-5A}) at an initial time $t=s$.
Let  $\varphi(\vec x,t,\hbar)$ be a function obtained from  Eq.
(\ref{borisov:LEQ-3}) and $\Psi (\vec x,t,\hbar)$ be the solution
of the nonlinear equation (\ref{borisov:GPE}) related to
$\varphi(\vec x,t,\hbar)$ according to (\ref{borisov:LEQ-3}).

Consider the operator $\hat A$, $\hat A:\mathcal K^s_\hbar \to$
$\mathcal K^s_\hbar$ such that
\[
[\hat A,\hat a_n(s)]=0,
\]
and a function $\varphi_A(\vec x,\hbar)=$ $\hat A\varphi(\vec
x,\hbar)\in{\mathcal K}^s_\hbar$. Then the function $\Psi_A (\vec
x,t,\hbar)$ related to $\varphi_A(\vec x,\hbar)$ by
(\ref{borisov:LEQ-3}) can be treated as a result of the action of
the symmetry operator $\hat A_{\rm nl}$  of the nonlinear equation
(\ref{borisov:GPE})
\[
\Psi_A (\vec x,t,\hbar)= \hat A_{\rm nl}\Psi (\vec x,t,\hbar).
\]

In conclusion, we note  that for a nonlinear Schr\"odinger
equation with a focusing nonlinearity, the many-dimensional
solutions localized at the initial time are unstable. This leads
to the phenomenon of collapse in the course of  evolution. The
semiclassical asymptotics (\ref{borisov:SOLUT-6}) behave in a
similar manner. They  can be constructed for special external
fields within  finite time intervals where  singularities typical
of collapse appear.

\section{ The three-dimensional anisotropic oscillator}

In  case of a harmonic oscillator field, the linear operator
$\hat{\mathcal H}$, equation (\ref{borisov:GPE1}), reads
\begin{gather}
\label{borisov:1+1GPE}  {\mathcal H}_0(\hat {\vec p},\vec
x,t)=\frac{\hat{\vec p}\,^2}{2m} +\frac{1}{2}\langle\vec x,K\vec
x\rangle,\qquad \vec x\in{\mathbb R}^3,
\end{gather}
where $K=\mathop{\rm diag}\{k_1,k_2,k_3\}$ with $k_j\in {\mathbb
R}^1$, $j=\overline{1,3}$.
 The  GPE  (\ref{borisov:GPE}) then takes the form
\begin{gather}
 \left[-i\hbar\partial_t+\frac{\hat{\vec p}\,^2}{2m}
+\frac{1}{2}\langle\vec x,K\vec x\rangle - g^{2}|\Psi
(x,t,\hbar)|^{2}\right]\Psi (x,t,\hbar)=0. \label{borisov:NESH}
\end{gather}
To construct the asymptotic solutions  (\ref{borisov:LEADT}) for
equation (\ref{borisov:NESH}) we solve the dynamic system
(\ref{borisov:HAMILT}) and  (\ref{borisov:SYSVAR}), which is
reduced to
\begin{gather}
  \dot{\vec p}=-K \vec x  +\frac{Q_2}{m}\vec\pi,
\qquad  \dot{\vec x}=\frac1m\vec p, \qquad
 \dot{\vec \pi}=-\frac{Q_1}{m}\vec\pi,
\label{borisov:system111}
\\
 \dot{\vec Z}= \frac1m\vec W, \qquad
\dot{\vec W}= -K\vec Z \label{borisov:system1}
\end{gather}
in the case under consideration. Here $\vec W_j(t)$ and $\vec
Z_j(t)$, $j=\overline{1,3}$,  are the linear independent solutions
of  the system in variations (\ref{borisov:system1}), which
determine the matrices $B(t)=\big(\vec W_1(t),\vec W_2(t),\vec
W_3(t)\big)$ and $C(t)= \big(\vec Z_1(t),\vec Z_2(t),$ $\vec
Z_3(t)\big)$; $Q(t)=Q_1(t)+iQ_2(t)=B(t)C^{-1}(t)$, the matrices
$Q_1(t)$ and  $Q_2(t)$ being real. The solution of  system
(\ref{borisov:system111}) and (\ref{borisov:system1}) is
substituted in (\ref{borisov:SOLUT-3a}), (\ref{borisov:SOLUT-4}),
(\ref{borisov:SOLUT-7}),  and (\ref{borisov:SOLUT-8}), and the
functions $S_\nu(\vec x,t,\hbar)$, $\sigma_\nu(\vec x,t,\hbar)$
are obtained. The  function   $\Psi_\nu (\vec x,t,\hbar)$ is then
determined by  (\ref{borisov:SOLUT-6}).

To write down the  solution of  systems (\ref{borisov:system111})
and (\ref{borisov:system1}) and the function $\Psi_\nu (\vec
x,t,\hbar)$, we introduce the notation $\Omega_j^+=\sqrt{k_j/m}$
for $k_j>0$, $\Omega_j^-=\sqrt{-k_j/m}$ for $k_j<0$,
$a_j(t)=\big(\vec W_j(t),\vec Z_j(t)\big)^\intercal$, and
$\{a_j(t), a_k(t)\}$ denoting the  skew-scalar product  of the
$6$-vectors $a_j(t)$, $a_k(t)$, $j,k=\overline{1,3}$.

Let us also introduce the functions
\begin{gather*}
S_{n_2,n_3}(t,\vec x,\vec p,\Omega_1,\Omega_2,\Omega_3,{\rm
z}(t))=S^0(t)+\displaystyle\frac m2\Omega_1 x^2_1 \mathop{\rm z}(
t)\\
\qquad {}+\frac{\pi
\hbar}{4}(n_2+n_3)-\hbar\left[\left(n_2+\frac12\right)\Omega_2+
\left(n_3+\frac12\right)\Omega_3\right]t+\langle \vec x,\vec
p\rangle, \\
 \sigma_{n_2,n_3}(\pi_0,t,\vec
x,\Omega_2,\Omega_3,{\rm \tilde{z}}(t))= \frac{\pi_0
x_1}{\mathop{\rm \tilde{z}}(t)}+ \frac{m\Omega_2}{2}
x_2^2+ \frac{m\Omega_3}{2} x_3^2-
\frac\hbar2\ln\mathop{|\rm
\tilde{z}}( t)| \\
\qquad {}-\hbar\ln\left[\frac1{\sqrt{n_2!n_3!}}
\sqrt[4]{\frac{m^2\Omega_2\Omega_3}{\pi^2\hbar^2}}
\left(\frac{1}{\sqrt{2}}
\right)^{n_2+n_3}H_{n_2}\left(\sqrt{\frac{m\Omega_2}
{\hbar}}x_2\right)H_{n_3}\left(\sqrt{\frac{m\Omega_3}
{\hbar}}x_3\right)\right].
\end{gather*}
Here  $\vec x=(x_1,x_2,x_3)$; $\vec p=(p_1,p_2,p_3)$ and $t$ are
real variables; $\pi_0$, $\Omega_1$, $\Omega_2$, $\Omega_3$ are
real constants; $\nu=(n_2, n_3)\in{\mathbb Z}_+^2$; ${\rm z}(t)$,
${\rm \tilde{z}}(t)$ are auxiliary real functions, whose form is
given below in each case, and  the function $S^0(t)$ is given by
\begin{gather}\label{borisov:S0}
 S^0(t)=\int^t_0\left[ \langle\dot{\vec
X}(t),\vec P(t)\rangle -\frac{\vec P^2(t)}{2m}
-\frac{1}{2}\langle\vec X(t),K\vec X(t)\rangle\right]dt.
\end{gather}
For the expressions  below to have a more compact form, we  also
use the following notation for trigonometric and hyperbolic
functions
\begin{gather*}
{\rm z^+}(t)=-{\rm tg}(t), \qquad {\rm z^-}(t)={\rm th}(t);
\qquad
{\rm \tilde{z}^+}(t)=\cos(t), \qquad {\rm \tilde{z}^-}(t)={\rm ch}(t); \\
{\rm \check{z}^+}(t)=\sin (t), \qquad {\rm \check{z}^-}(t)={\rm sh}(t).
\end{gather*}
The solutions of  system in variations, equation
(\ref{borisov:system1}), are normalized by the conditions
\begin{gather}
\{a_j(t), a_k(t)\}=\{a_j^*(t), a_k^*(t)\}=0, \qquad \{a_j(t),
a_k^*(t)\}= 2i \big(\delta_{kj}-\delta_{1k}\delta_{1j}\big),\quad
k,j=\overline{1,3}. \label{borisov:matr1}
\end{gather}
The  solutions of  system  (\ref{borisov:system1}) normalized by
condition (\ref{borisov:matr1}) determine the  matrices
$Q_1^\pm(t)$ and  $Q_2^\pm(t)$, which are found as
\[
Q_1^\pm(t)=\mathop{\rm diag}\,\{\mp m\Omega_1^\pm \mathop{\rm
z^\pm}[\Omega_1^\pm t],0,0\}, \qquad Q_2^\pm(t)=\mathop{\rm
diag}\,\{0,m\Omega_2^\pm ,m\Omega_3^\pm\}.
\]

The vector $\vec \pi (t)$ then  takes the form
\[
\vec \pi{}^\pm(t)= \pi_0^\pm\left(\frac1{\mathop{\rm
\tilde{z}^\pm}[\Omega_1^\pm t]},0,0\right)^\intercal ,
\]
where  $ \pi_0^\pm$ are  constants of integration.

For the case considered  we have $\dot{\vec p}{}^\pm=-K \vec x
{}^\pm$, $\dot{\vec x}{}^\pm=\frac1m\vec p {}^\pm,$ and
\[
\vec P^\pm(t)=\big({\mp\check R^\pm\Omega_1^\pm\mathop{\rm
\check{z}^\pm}[\Omega_1^\pm t+\varphi_0^\pm]},0,0\big)^\intercal,
\qquad \vec X^\pm(t)=\big({\tilde R^\pm\mathop{\rm
\tilde{z}^\pm}[\Omega_1^\pm
t+\varphi_0^\pm]}/m,0,0\big)^\intercal,
\]
where  $\check R^\pm$ and $\tilde R^\pm$ are constants of
integration. Then, following (\ref{borisov:SOLUT-6}),
(\ref{borisov:SOLUT-7}), and (\ref{borisov:SOLUT-8}), we obtain
the functions
\begin{gather}
 \label{borisov:matr11f} \Psi^\pm_{n_2,n_3}(\vec
x,t,\hbar)=\frac
{|\vec\pi^\pm(t)|}{\sqrt{m}g}\frac{\exp\left[\frac{i}{\hbar}
S_{n_2,n_3}\big(t,\Delta\vec x^\pm,\vec
P^\pm(t),\Omega_1^\pm,\Omega_2^\pm,\Omega_3^\pm,{\rm
z^\pm}(\Omega_1^\pm t)\big)\right]} {\mathop{\rm
ch}\left[\frac{1}{\hbar}
\sigma_{n_2,n_3}\big(\pi_0^\pm,t,\vec{\Delta x^\pm},\Omega_1^\pm,
\Omega_2^\pm,\Omega_3^\pm,{\rm \tilde{z}^\pm}(\Omega_1^\pm
t)\big)\right]}.
\end{gather}
Here the multiindex $\nu$ in  general formula
(\ref{borisov:SOLUT-6}) is of the form $\nu=(n_2, n_3)\in{\mathbb
Z}_+^2$.

The  functions (\ref{borisov:matr11f}) can be considered excited
states for equation (\ref{borisov:GPE}) with equation
(\ref{borisov:LASE}) being the linear associated Schr\"odinger
equation.

Note that for $n_2=n_3=0$, the expression (\ref{borisov:matr11f})
gives the ``vacuum'' solution
 (\ref{borisov:VACUUM4}).

Let $\varphi(x_2,x_3)$ be a function of
 ${\mathcal S} ^t_\hbar$ class. Consider the Cauchy problem for
equation (\ref{borisov:NESH})
\[
 \Psi\Big|_{t=s}=\frac
{|\pi_0|}{\sqrt{m}g}\frac{\exp\big( i\,\mathop{\rm
Im}\,\ln[\varphi(x_2,x_3)]\big)}{\mathop{\rm
ch}\left(\frac{1}{\hbar} \pi_0(x_1-x_{10})+\mathop{\rm
Re}\ln[\varphi(x_2,x_3)]\right)}.
\]
Let  $\hat U^{(0)}_{\mathcal H_0}(t)$ denote the linear evolution
operator
\[
\varphi(x_2,x_3,t)=\hat U^{(0)}_{\mathcal
H_0}(t,s)\varphi(x_2,x_3)=\int_{\mathbb R^2}G^{(0)}_{\mathcal
H_0}(x_2,x_3,y_2,y_3,t,s) \varphi(y_2,y_3)dy_2dy_3
\]
with the kernel
\begin{gather*}
G^{(0)}_{{\mathcal H}_0}
(x_2,x_3,y_2,y_3,t,s)=\sum^{\infty}_{n_2=n_3=0}
\frac1{n_2!n_3!}\left(\frac{1}{{2}}\right)^{n_2+n_3}H_{n_2}\left(\sqrt{\frac{m\Omega_2}
{\hbar}}\Delta
x_2\right) \\
\qquad{}\times H_{n_3}\left(\sqrt{\frac{m\Omega_3}{\hbar}}\Delta
x_3\right)
H_{n_2}\left(\sqrt{\frac{m\Omega_2} {\hbar}}\Delta
y_2\right)H_{n_3}\left(\sqrt{\frac{m\Omega_3} {\hbar}}\Delta
y_3\right)\\
\qquad {}\times \exp\left\{-i\left[\left(n_2+\frac12\right)\Omega_2+\left(n_3+\frac12\right)\Omega_3\right](t-s)\right\}\\
\qquad{}\times \exp\left[-\frac{m} {2\hbar}\left(\Omega_2\big(\Delta
x_2^2+\Delta y_2^2\big)+\Omega_3\big(\Delta x^2_3+\Delta
y_3^2\big)\right)\right]\sqrt{\frac{m^2\Omega_2\Omega_3}{\pi^2\hbar^2}},\\
 \Delta y_j=y_j-x_{j0}, \qquad \Delta x_j=x_j-X_{j0}(t), \qquad j=2,3.
\end{gather*}
Using the Mehler formula
\[
\sum_{n=0}^\infty\frac
1{n!}\left(\frac\lambda2\right)^nH_n(x)H_n(y)=
\frac1{\sqrt{1-\lambda^2}}\exp\left[\frac{2xy\lambda-(x^2+y^2)\lambda^2}
{1-\lambda^2}\right],
\]
where $\lambda$ is an arbitrary complex
parameter, $|\lambda|\leq 1$, we obtain
\begin{gather*}
G^{(0)}_{{\mathcal H}_0}(x_2,x_3,y_2,y_3,t,s)\\
\qquad {}= \sqrt{g_\hbar(t-s)}\exp\left[\frac{i}{\hbar}\big(f(t,s,\Omega_2,\Delta
x_2,\Delta y_2
)+f(t,s,\Omega_3,\Delta x_3,\Delta
y_3)\big)\right].
\end{gather*}
Here
\begin{gather*}
g_\hbar(t-s)=-\frac{m^2\Omega_2\Omega_3}{4\pi^2 \hbar^2 \sin (\Omega_2(t-s)) \sin (\Omega_3(t-s))},\\
f(t,s,\Omega,x,y)=-\frac{m\Omega\big(2
xy-\big(x^2+y^2\big)\cos(\Omega (t-s))\big)}{2\sin(\Omega (t-s))}.
\end{gather*}

Define the operator $\hat U^{(0)}_{\rm tr}$ by its action on the
function $\varphi (x_2,x_3)\in {\mathcal S} ^t_\hbar$ as
\begin{gather}
 \label{borisov:matr12p} \Psi^\pm (\vec
x,t,\hbar)=\hat U^{(0)\pm}_{\rm tr}(\varphi)(\vec x,t)= \frac
{|\pi^\pm_0|}{\sqrt{m}g|{\rm \tilde{z}^\pm}(\Omega_1^\pm t)|}
\exp\left(\frac{i}{\hbar} S^\pm_{\varphi}(\vec x,t)\right)
\mathop{\rm ch}{}^{-1}\left(\frac{1}{\hbar}
\sigma^\pm_{\varphi}(\vec x,t)\right),
\end{gather}
where
\begin{gather}
S^\pm_{\varphi}(\vec x,t)={S^{0}}^\pm(t)+P^\pm_1(t)\Delta
x^\pm_1+\frac12m\Omega^\pm_1(\Delta x^\pm_1)^2 \mathop{\rm
z^\pm}[\Omega_1^\pm t]
-\hbar\, {\rm Im}\left(\ln[\varphi(x_2,x_3,t)]\right), \label{borisov:matr12pa}  \\
\sigma^\pm_{\varphi}(\vec x,t)=\frac{\pi^\pm_0\Delta x^\pm_1}{{\rm
\tilde{z}^\pm}[\Omega_1^\pm t]}- \frac\hbar2\ln|{\rm
\tilde{z}^\pm}[\Omega_1^\pm t]| +\hbar\,{\rm
Re}\left(\ln[\varphi(x_2,x_3,t)]\right). \label{borisov:matr12paa}
\end{gather}
Here ${S^0}^\pm$ is given by  (\ref{borisov:S0}) with
$\vec{P}(t)=\vec{P}(t)^\pm$ and $\vec{X}(t)=\vec{X}(t)^\pm$,
respectively.

We call  the operator $\hat U^{(0)}_{\rm tr}$ of the form
(\ref{borisov:matr12p}), (\ref{borisov:matr12pa}), and
(\ref{borisov:matr12paa}) the {\it  ``transverse''  evolution
operator} for equation (\ref{borisov:NESH}). To construct
localized solutions for equation (\ref{borisov:NESH}), we can
take, for example, the function
\[
\varphi(x_2,x_3)=\exp\left[-\frac1\hbar\left( x_2^2+
x_3^2\right)\right]
\]
as the initial function, i.e., at $t=0$.

Below we put  $m=1$, $g=1$, $K=E$ for $\Psi^+=\Psi^+_{00}$, and $K=-E$ for
$\Psi^-=\Psi^-_{00}$, where  $E$ is the unit matrix and $\Psi^{\pm}_{00}$
is the ``vacuum'' function (\ref{borisov:matr11f}).

Then, for the $(1+1)$-dimensional case, the module of  function
(\ref{borisov:matr11f}) under  the initial conditions $P(0) = 0$,
$x(0) = 0$, $\pi_0=1$, $\sigma _{1}(0)=0$ take the form
\begin{gather}
 |\Psi^+(x,t,\hbar)|=|\cos(t)|^{-1}\mathop{\rm
ch}{}^{-1}\left[\frac{1}{\hbar}\left(\frac{x}{\cos(t)}-
\frac\hbar2\ln|\cos(t)|\right)\right] \qquad {\mathop{\rm
for}}\qquad  K=k=1\label{borisov:well}
\end{gather}
or
\begin{gather}
 |\Psi^-(x,t,\hbar)|=|\mathop{\rm
ch}(t)|^{-1}\mathop{\rm
ch}{}^{-1}\left[\frac{1}{\hbar}\left(\frac{x}{\mathop{\rm ch}(t)}-
\frac\hbar2\ln\mathop{\rm ch}(t)\right)\right] \qquad {\mathop{\rm
for}}\qquad K=k=-1.\label{borisov:hill}
\end{gather}

\begin{figure}[t]
\centerline{\includegraphics[width=6cm]{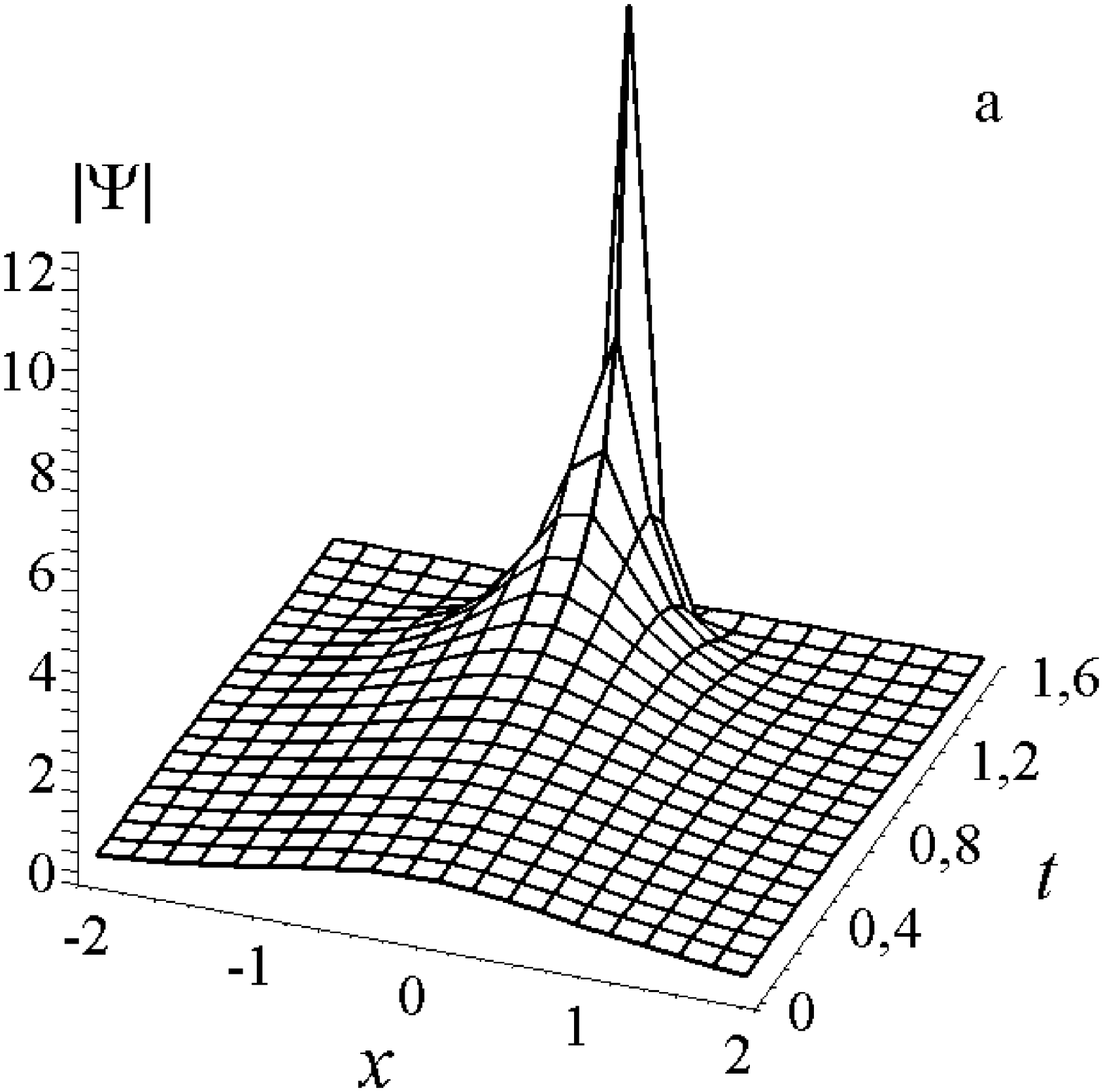}\qquad \includegraphics[width=6cm]{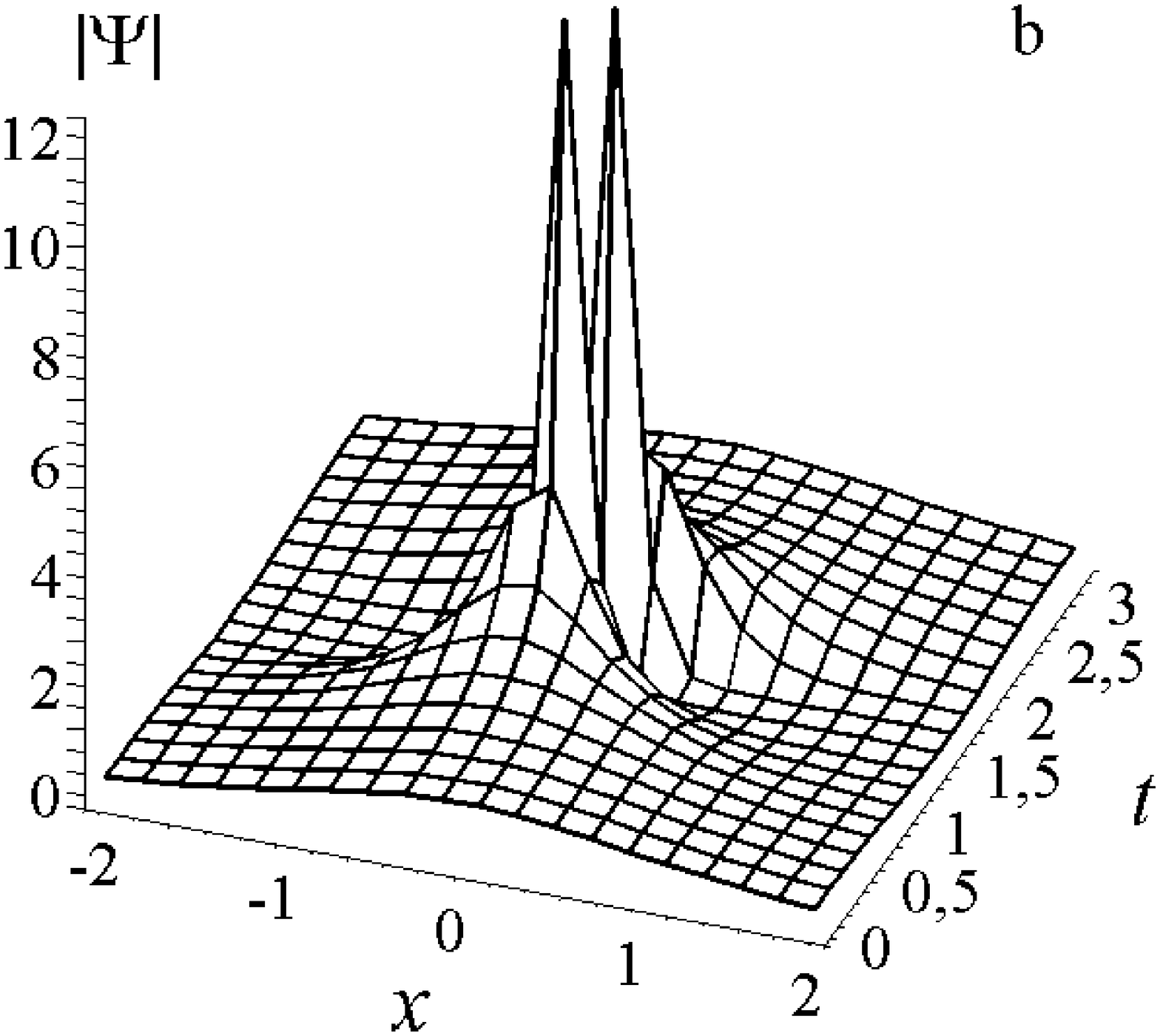}}
\vspace{-3mm}

\caption{} \label{borisov:Fig2}
\end{figure}

In Fig.~\ref{borisov:Fig2}a, the function $|\Psi^+|$
(\ref{borisov:well}) is shown  for the $(1+1)$  space-time  and
the  potential well (${kx^2}/{2}$, $k>0$) of the oscillator form
in (\ref{borisov:1+1GPE}) with $\hbar=1$. It can be seen that the
solution, being localized at the initial  time, is focused  within
the time interval $(0,\pi/2)$ (in the process of evolution) and
collapses. The GPE with an external field was used  to describe
the dispersion and diffraction of nonlinear waves
\cite{borisov:BANG}. The collapse phenomenon was studied in
\cite{borisov:BANG}  by   numerical simulations. The numerical
results qualitatively correlate with those obtained by analytic
asymptotic methods. The asymptotic solution
(\ref{borisov:matr12p}) describes the system for any $T$ within
a~finite time interval $[0, T]$. If the time interval is short
enough, the evolution can end before the collapse occurs.

\begin{figure}[t]
\centerline{\includegraphics[width=6cm]{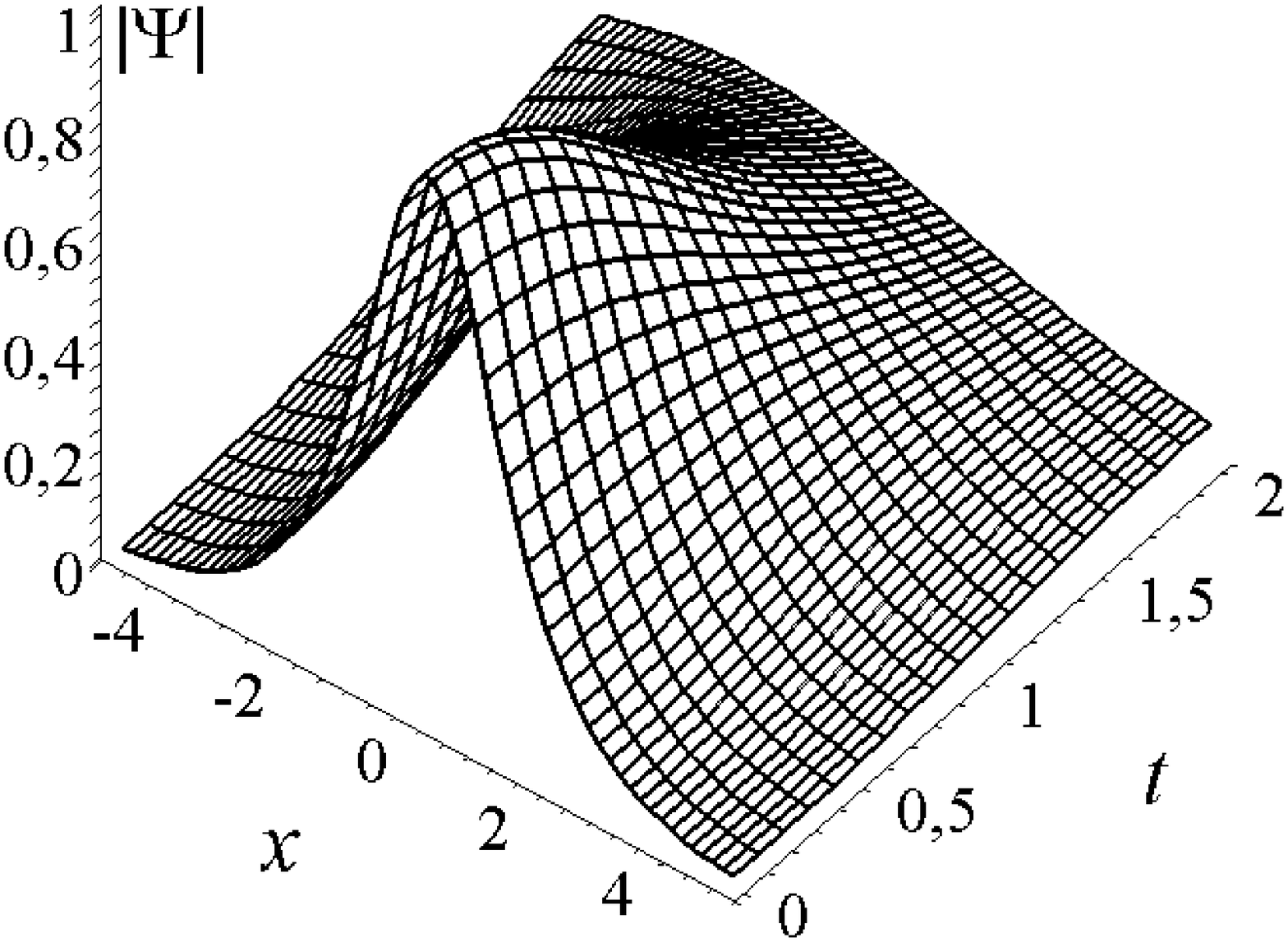}}

\vspace{-3mm}

\caption{} \label{borisov:Fig3}
\end{figure}

The function (\ref{borisov:well}) is $\pi$-periodic and its graph
is given in Fig.~\ref{borisov:Fig2}b. The term
$(-\hbar/2)\ln|\cos(t)|$ in the argument of hyperbolic cosine in
equation~(\ref{borisov:well}) shifts the function centroid
(\ref{borisov:well}) within the time interval $0<t<{\pi}/{2}$.
Throughout  the evolution time, this gives rise to oscillations of
the function maximum about the plane $x=0$.

For the potential hill of the oscillator form (${kx^2}/{2}$,
$k<0$) in equation~(\ref{borisov:1+1GPE}),  the state of the
system  in the $(1+1)$ space-time  is described by $\Psi^-$. Let
us choose the initial condition for the function $\Psi^-$  the
same as for $\Psi^+$. Then the dynamics of the system is
characterized by defocusing and  exponential damping (see
Fig.~\ref{borisov:Fig3}). The term $(-\hbar/2)\ln(\mathop{\rm
ch}(t))$ in the argument of hyperbolic cosine in
(\ref{borisov:hill})  shifts the function centroid in the negative
direction of $x$.

\begin{figure}[t]
\centerline{\includegraphics[width=5cm]{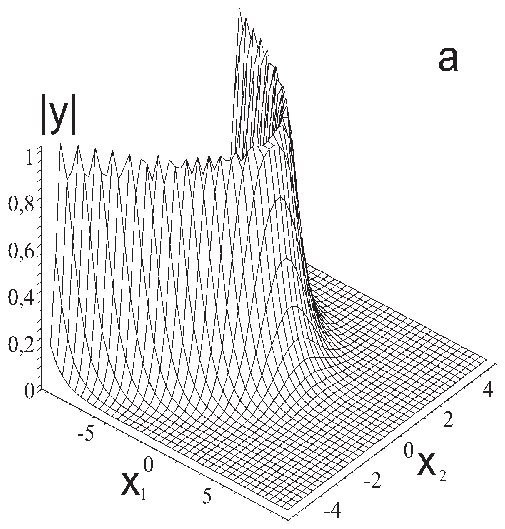}\quad
\includegraphics[width=5cm]{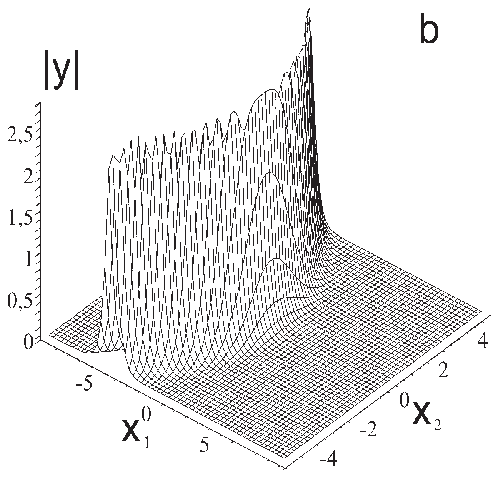}\quad
\includegraphics[width=5cm]{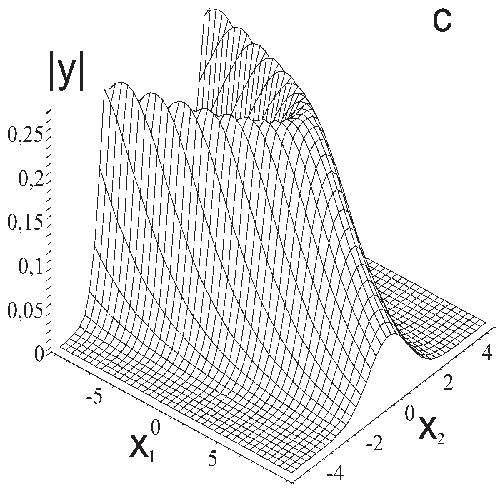}}
\vspace{-3mm}

\caption{} \label{borisov:Fig4}
\end{figure}

Consider the $(2+1)$-dimensional case. For the functions
(\ref{borisov:matr11f})  put \mbox{$\vec P(0) = 0$}, $\vec x(0) =
0$, $\pi_0=1$, $\sigma _{1}(0)=0$, then we have
\begin{gather}
|\Psi^+(\vec x,t,\hbar)|=|\cos(t)|^{-1}\mathop{\rm
ch}{}^{-1}
\left[\frac{1}{\hbar}\left(\frac{x_1}{\cos(t)}+\frac{x_2^2}{2}-
\frac\hbar2\ln|\cos(t)|\right)\right]\label{borisov:well21},
\\
\displaystyle  |\Psi^-(\vec x,t,\hbar)|=|\mathop{\rm
ch}(t)|^{-1}\mathop{\rm ch}{}^{-1}
\left[\frac{1}{\hbar}\left(\frac{x_1}{\mathop{\rm
ch}(t)}+\frac{x_2^2}{2}- \frac\hbar2\ln\mathop{\rm
ch}(t)\right)\right]\label{borisov:hill21}.
\end{gather}
The  functions (\ref{borisov:well21}) and (\ref{borisov:hill21})
coincide at $t=0$.  The initial graph of  the functions
(\ref{borisov:well21}) and~(\ref{borisov:hill21}) is presented in
Fig.~\ref{borisov:Fig4}a.  Figs.~\ref{borisov:Fig4}b,c show the
graphs  of  function~(\ref{borisov:well21})  for  $t=1.2$ and of
the function~(\ref{borisov:hill21}) for  $t=2$, respectively.

The solution considered  is localized about the parabola with the
vertex at the  point  $\vec {P}\left( {t} \right) = 0$, $\vec
{X}\left( {t} \right) = 0$ of  phase space. In the course of
evolution of  $|\Psi^+|$, the parabola branches disperse and at
$t=\pi/2$ it  transforms   into a straight line. The inverse
behavior of the parabola takes place for the
 evolution of $|\Psi^-|$: the parabola branches converge for an infinite time.
The amplitude of the functions $|\Psi^{\pm}|$ behaves, in the
direction along the coordinate $x$, as in the $(1+1)$-dimensional
case.

\section{Conclusion remarks}

The  asymptotics obtained, equation (\ref{borisov:LEADT}), can be
regarded as a necessary step in the construction of a global
semiclassical
 asymptotic solution to the GPE (\ref{borisov:GPE}).
It may be supposed that   the  functions (\ref{borisov:LEADT})
describe the behavior of an element of the global solution in the
neighborhood of a ray along the normal $\vec\pi$
(Fig.~\ref{borisov:Fig1}) to the (closed) surface about which the
global solution is concentrated. Substantiation of  these global
asymptotics for finite times $t\in[0,T]$, $T=\mbox{\rm const}$, is
a special nontrivial mathematical problem. This problem is
concerned with obtaining {\em a priori} estimates for the solution
of a nonlinear equation, which are uniform in parameter
$\hbar\in]0,1]$ and is beyond the scope of the present work. Note
that, in view of the heuristic considerations given in
\cite{borisov:Mas3}, it seems that the difference between the
exact and the constructed formal asymptotic solution can be found
with the use of method developed in \cite{borisov:Mas3,
borisov:Mas3a}.

The technique of construction of semiclassical asymptotics
developed in Section~\ref{borisov:SecGPE} provides a way for
solving the  problem of correspondence between the classical and
quantum results for quantum systems described by nonlinear
equations, namely, via finding solutions to the dynamic system
(\ref{borisov:HAMILT}) and (\ref{borisov:SYSVAR}). For nonlinear
systems this problem  differs from the relevant problem in the
linear quantum mechanics. For linear quantum systems, a correct
formal transition from the quantum theory to the classical one
requires imposing  special limitations  on the quantum states
(their semiclassical concentration). The states not satisfying
these limitations are regarded as ``essentially quantum'', and
those satisfying them are considered ``near to classical''. Thus,
the dynamics of the classical objects obtained is described by the
classical Hamiltonian equations no matter the domain where the
wave function is concentrated (whether it be a point, a curve, or
a surface). The study of the dynamic system (\ref{borisov:HAMILT})
and  (\ref{borisov:SYSVAR}) is a separate  mathematical subject
for research. For example, the Hamiltonian or Poissonian
formalisms as applied to this system are of interest.

The construction of  ``transverse'' evolution operator given in
Section~\ref{borisov:SecGPE} allows one not only to obtain
semiclassical asymptotics but also to construct approximate
symmetry operators of special type acting in the class of
${\mathcal S}_\hbar^t$ functions under consideration. Such
operators can be naturally referred to as semiclassical symmetry
operators~\cite{borisov:Shv02} (see also~\cite{borisov:LTS}).

\subsection*{Acknowledgements}
 The work was supported in part by a Grant of President of the
Russian Federation (No.\ NSh-1743.2003.2).

\LastPageEnding

\end{document}